\documentclass[twocolumn, twocolappendix]{aastex631}
\usepackage{bm}
\usepackage{CJKutf8}
\usepackage{amsmath}
\usepackage{graphicx}

\shorttitle{Bifurcated evolutionary path}
\shortauthors{Tao Fu et al.}
\graphicspath{{./}{figures/}}

\begin{document}
\begin{CJK}{UTF8}{gbsn}

\title{Bifurcated Evolutionary Pathways in Multi-planet Systems Driven by Misaligned Protoplanetary Disks}

\correspondingauthor{Yue Wang}
\email{ywang@buaa.edu.cn}

\author[0000-0002-9981-0761]{Tao Fu (伏韬)}
\affiliation{School of Astronautics, Beihang University, Beijing 102206, People's Republic of China}

\author[0000-0001-9172-0895]{Yue Wang (王悦)}
\affiliation{School of Astronautics, Beihang University, Beijing 102206, People's Republic of China}


\author[0000-0003-0090-3675]{Weiduo Hu (胡维多)}
\affiliation{School of Astronautics, Beihang University, Beijing 102206, People's Republic of China}



 \begin{abstract}

 Stellar obliquities, or spin-orbit angles, prevalent in exoplanet systems, can impose important constraints on their formation and evolution histories. Recent studies suggest that primordial misalignments between protoplanetary disks and stellar spin axes may significantly contribute to these obliquities, as those frequently observed in systems hosting hot Jupiters. In this study, we demonstrate that misaligned protoplanetary disks combined with stellar oblateness drive complex dynamical evolution in planetary systems during their disk dispersal stages. Specifically, we identify bifurcated evolutionary pathways in multi-planet systems: systems with low star-disk misalignment angles ($\psi_{\star0}$) undergo smooth, adiabatic evolution, producing nearly coplanar, low-obliquity configurations; in contrast, systems with high misalignment angles typically experience an abrupt, non-adiabatic transition, leading to large-amplitude libration of mutual planetary inclinations and then triggering chaotic eccentricity excitation. This libration and eccentricity excitation process can propagate inward-outward in compact multi-planet systems, forming an excitation chain that can destabilize the entire system. The non-adiabatic transition arises from the dynamical bifurcation-induced effect, which occurs during disk dissipation when $\psi_{\star0}\gtrsim44.6^\circ$ (for one-planet systems). Our framework predicts that surviving typical compact multi-planet systems originating from misaligned disks evolve toward coplanar, low-obliquity configurations, consistent with observations of \textit{Kepler} multi-planet systems. These results advance our understanding of planetary dynamics in misaligned disks and their evolutionary outcomes.
 \end{abstract}



\section{Introduction} \label{sec:intro}

 Unlike the well-aligned configuration of our solar system, exoplanets are frequently observed on highly inclined orbits relative to their host stars' equators, which is one of the most striking discoveries in exoplanet research over the past decades \citep{schlaufman2010evidence, winn2010hot}. Stellar obliquity, defined as the angle between a planet's orbital angular momentum vector and its host star’s spin axis, as typically measured via the Rossiter-McLaughlin (RM) effect \citep{mclaughlin1924some,rossiter1924detection}, serves as a critical constraint on the formation and dynamical evolution of planetary systems \citep{fabrycky2007shrinking}. A prominent example lies in the obliquity distribution of hot stars hosting hot Jupiters (HJs): these systems span a wide range of spin-orbit angles ($0\text{-}180^\circ$) \citep{albrecht2012obliquities}, which have been regarded as critical clues for deciphering the origin and evolution of HJs \citep{dawson2018origins}.

 Two primary mechanisms have been proposed to explain HJs' high obliquities associated with their formation channels. The first involves the violent high-eccentricity tidal migration through post-disk multi-body interactions.  In contrast, the second mechanism posits primordial star-disk misalignments, which could enable in situ formation or disk-aligned migration of HJs \citep{batygin2012primordial,batygin2016situ,boley2016situ}. By synthesizing related evidence, recent studies suggest that the two origin channels together are responsible for the observed population of HJs: high-eccentricity tidal migration primarily triggered by planet-planet ZLK oscillations dominates as the predominant channel, while in situ formation or disk migration supplements this process as a secondary pathway \citep{dawson2018origins}.

 However, recent studies have identified critical limitations in invoking the planet-planet ZLK effect as the dominant mechanism for HJ formation. Key evidence includes: (1) the initial high planetary inclinations required for triggering the ZLK effect should be rare \citep{becker2017exterior}; (2) the absence of super-eccentric proto-hot Jupiters in the \textit{Kepler} sample undermines predictions of high-eccentricity tidal migration; (3) HJs are not as lonely as previously thought, as more than $12\% \pm 6\%$ HJs have nearby planetary companions \citep{wu2023evidence}; (4) a broad, zero-centered unimodal obliquity distribution identified through hierarchical Bayesian analysis disfavors the multimodal obliquity patterns expected from ZLK-driven processes \citep{dong2023hierarchical}. 

 On the other hand, analyses of  planetary systems with giant planets and HJs with giant companions suggest that HJs' observed properties align more closely with coplanar high-eccentricity tidal migration as the dominant formation channel \citep{zink2023hot}. Notably, this channel requires additional processes, typically primordial misalignments, to produce the observed high stellar obliquities. Consequently, the primordial star-disk misalignment is believed to have played an important role in HJ formation.

 Direct observations of protoplanetary disks further support the prevalence of primordial star-disk misalignments, though current samples remain limited due to the challenges of resolving disk orientations and stellar spin axes simultaneously. For instance, \citet{davies2019star} identified misalignments in $\sim 33\%$ (5 out of 15) of young star-disk systems by comparing inclinations of stellar and disk angular momentum vectors, while \citet{hurt2023evidence} reported $\sim19\%$ (6 out of 31) misalignment in debris disk systems. Individual evidence include the recent discovery of a highly misaligned protoplanetary disk around a 3-Myr protostar \citep{barber2024giant}. Additionally, moderate primordial misalignments are theorized to enable the formation of polar-aligned circumbinary disks, which were previously predicted to be exist and have been confirmed by recent observations \citep{kennedy2019circumbinary}.

 Recent studies leveraging primordial star-disk misalignment have successfully explained several critical phenomena in exoplanetary systems. For example, a tilted oblate star could drive moderate mutual planetary inclinations, addressing the phenomenon of \textit{Kepler} dichotomy \citep{spalding2016spin,dai2018larger}. The high stellar obliquity of the two coplanar planets ($\psi_{\star}=124^\circ \pm 6^\circ$) in the K2-290 A system has been attributed to a primordial star-disk misalignment caused by the gravitational torque from its stellar companion K2-290 B \citep{hjorth2021backward}. Furthermore, \citet{vick2023high} demonstrated that high-eccentricity tidal migration with primordial disk misalignments (induced by stellar companions) closely reproduces the observed subpopulation of HJs with orbits perpendicular to their host stars' equatorial planes \citep{albrecht2021preponderance}.

 Particularly, \citet{fu2024sculpting} have recently identified a novel mechanism driven by the combined effects of the full-space gravity of a disk and stellar oblateness during disk dispersal, operating at polar primordial misalignment angles ($\psi_{\star 0}>70^\circ$). Based on this mechanism, they provided a plausible explanation for the peculiar near-perpendicular, retrograde orbital configuration in the HD 3167 system. In this paper, we generalize this effect, termed the dynamical bifurcation-induced effect, to the entire spectrum of primordial stellar obliquities. Crucially, we demonstrate that this mechanism has profound consequences for multi-planet systems, producing bifurcated evolutionary pathways primarily depending on the extent of primordial misalignments.

 Previous studies focusing on planet-disk interactions failed to identify this critical mechanism. They either adopted an approximate disk potential that is only feasible when planets are close to the disk plane, thus ignoring critical nonlinear effects \citep{spalding2020stellar, zanazzi2018planet}; or they assumed the inner region of the disk had already completely dissipated, causing the secular perturbation from the outer disk resemble that from an outer perturbing body \citep{millholland2019excitation,su2020dynamics,petrovich2020disk}. In contrast, our study is based on the gravitational potential of the disk that is feasible to the full physical space \citep{schulz2012gravitational,fu2024sculpting}.

 The structure of this paper is organized as follows. Sec. \ref{sec:DBI} introduces the dynamical-bifurcation-induced effect through a simplified system configuration, establishing its theoretical foundations. Sec. \ref{sec:BEP} explores how this mechanism drives bifurcated evolutionary pathways (adiabatic vs. non-adiabatic) in multi-planet systems. Sec. \ref{sec:CEC} analyzes the eccentricity excitation mechanism and the establishment of an excitation chain in the non-adiabatic evolution pathway. Sec. \ref{sec:SA} evaluates the prevalence of our framework through statistical analysis of \textit{Kepler} multi-planet systems. Sec. \ref{sec:SCMPS} presents available evidence supporting the proposed framework and examines the orbital architectures sculpted by it. Finally, we conclude in Sec. \ref{sec:conclusions}.

\section{Dynamical-bifurcation-induced effect} \label{sec:DBI}

 To develop a qualitative understanding of the fundamental mechanism driven by a misaligned disk, we first consider a simple scenario: a single close-in planet (with mass $m_1$ and semi-major axis $a_1$) initially embedded within a protoplanetary disk (oriented along $\hat{\bm{l}}_{\mathrm{d}}$) that is misaligned with the stellar spin axis (oriented along $\hat{\bm{s}}_{\star}$) by a misalignment angle $\psi _{\star 0}$ (also referred to as the primordial obliquity). The planet's orbital evolution is governed by the combined gravitational influence of the disk and the star's rotational oblateness. Initially, the planet resides in a circular orbit ($e_1=0$) due to the disk's damping effect. Since $e_1=0$ is a equilibrium, we first restrict our analysis to orbital precession by assuming $e_1=0$. Therefore, the secular potential governing the planet's precession is given by (Appendix \ref{app_EqOM})
\begin{equation}
	\Phi _{\mathrm{circ}}=-\frac{\phi _{\star,1}}{2}\left( \boldsymbol{l}_1\cdot \hat{\boldsymbol{s}}_{\star} \right) ^2+\phi _{\mathrm{dk},1}S_1\left( 1-\frac{\pi}{8}S_1 \right) 
\label{eq:1}
\end{equation}
Here $\boldsymbol{l}_1=\sqrt{1-e^2_1}\hat{\boldsymbol{l}}_1$ is the normalized orbit angular momentum vector, $S_1=\sqrt{1-\left( \hat{\boldsymbol{l}}_1\cdot \hat{\boldsymbol{l}}_{\mathrm{dk}} \right) ^2}=\sin{I_{1\rm{d}}}$, where $I_{1\rm{d}}$ denotes the orbit inclination relative to the disk plane,
\begin{equation}
	\phi _{\star, 1}=\frac{3Gm_{\star}m_1J_2R_{\star}^{2}}{2a_{1}^{3}} \,,
\label{eq:2}
\end{equation}
where $m_{\star}$ and $R_{\star}$ are the mass and radius of the central star, and $J_2$ is the coefficient of stellar oblateness, and
\begin{equation}
	\phi _{\mathrm{dk},1}=2\pi m_1 G\Sigma _{a_1}a_1\,.
\label{eq:3}
\end{equation}
where $\Sigma_{a_1}$ is the disk surface density at the radius $a_1$.

Then, by using the Lagrange planetary equations in terms of vectorial elements $\boldsymbol{l}_1$ and $\boldsymbol{e}_1$ (Eq. (\ref{eq:A20})), the precession equation is given by
\begin{equation}
	\dot{\boldsymbol{l}}_1=-\left( \boldsymbol{l}_1\cdot \hat{\boldsymbol{s}}_{\star} \right) \hat{\boldsymbol{s}}_{\star}\times \boldsymbol{l}_1-\frac{1}{\epsilon _{\star \mathrm{d}}}\frac{1}{S_1}\left( 1-\frac{\pi}{4}S_1 \right) \left( \hat{\boldsymbol{l}}_{\mathrm{d}}\cdot \hat{\boldsymbol{l}}_1 \right) \hat{\boldsymbol{l}}_{\mathrm{d}}\times \boldsymbol{l}_1
\label{eq:4}
\end{equation}
where we define
\begin{equation}
	\epsilon _{\star \mathrm{d}}=\frac{\phi _{\star,1}}{\phi _{\mathrm{dk},1}}=\frac{3}{2}\frac{m_{\star}}{2\pi \Sigma_{a_1}  a_{1}^{2}}\frac{J_2R_{\star}^{2}}{a_{1}^{2}},
\label{eq:5}
\end{equation}
which characterizes the relative strength between the stellar oblateness and the disk's gravity.

To avoid the singularity at $I_{1\rm{d}}=0$, we introduce a softening scale $\beta$ for the disk's gravity by replacing $S_1$ with $S_1=\sin I_{1\mathrm{d}}+\beta$. For the circular problem, there are two constants of motion , $\Phi_{\rm{circ}}$ and $\left| \boldsymbol{l}_1 \right|=1$, suggesting that the system is integrable.

The equilibrium states of the system can be derived from $\dot{\boldsymbol{l}}_1=0$. According to Eq. (\ref{eq:4}), there are two distinct types of equilibrium: one characterized by $\boldsymbol{l}_1$ lying within the plane defined by $\hat{\boldsymbol{s}}_{\star}$ and $\hat{\boldsymbol{l}}_{\mathrm{d}}$, referred to as the coplanar equilibrium; the other featured by $\boldsymbol{l}_1$ being perpendicular to $\hat{\boldsymbol{s}}_{\star}$ and $\hat{\boldsymbol{l}}_{\mathrm{d}}$, referred to as the orthogonal equilibrium.

For the coplanar equilibrium, the direction of the equilibrium can be specified by the azimuth angle of $\boldsymbol{l}_1$ relative to $\hat{\boldsymbol{s}}_{\star}$, defined by $\psi_{1\star}$. The azimuth angle relative to $\hat{\boldsymbol{l}}_{\mathrm{d}}$ is then $\psi _{1\star}-\psi _{\star 0}$. By using these definitions, the coplanar equilibrium conditions becomes
\begin{equation}
\begin{split}
    f\left( \psi _{1\star}; \psi _{\star 0}, \epsilon _{\star \mathrm{d}} \right) &\triangleq \sin 2\psi _{1\star}-\frac{1}{\epsilon _{\star \mathrm{d}}}\frac{1}{S_1}\left( 1-\frac{\pi}{4}S_1 \right) \\
    &~~~ \times \sin 2\left( \psi _{\star 0}-\psi _{1\star} \right) =0 \,,
\end{split}
\label{eq:6}
\end{equation}
where $S_1=\left| \sin \left( \psi _{\star 0}-\psi _{1\star} \right) \right|+\beta$.

The invariance of Eq. (\ref{eq:4}) under the transformations $\hat{\boldsymbol{s}}\rightarrow -\hat{\boldsymbol{s}}_{\star}$ or $\hat{\boldsymbol{l}}_{\mathrm{d}}\rightarrow -\hat{\boldsymbol{l}}_{\mathrm{d}}$ implies that we only need to consider the primordial obliquity ranging from 0 to $90^\circ$. Solutions for $\psi _{\star 0}\in \left( 90^\circ, 180^\circ \right) $ are trivially recovered via symmetry. In addition, invariance under the transformation $\left( \boldsymbol{l}_1, t \right) \rightarrow \left( -\boldsymbol{l}_1,-t \right) $ enforces symmetry in the phase space between $\boldsymbol{l}_1$ and $-\boldsymbol{l}_1$. Consequently, we restrict our analysis to $\psi _{1\star}\in \left( 0, 180^\circ \right) $.

\begin{figure}
\plotone{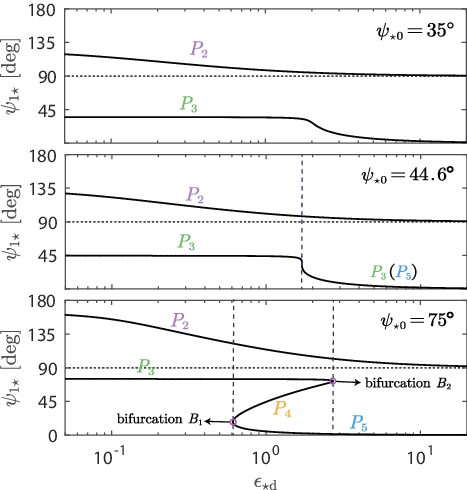}
\caption{Coplanar equilibrium solutions as a function of $\epsilon_{\star \rm{d}}$  for $\psi_{\star0}=35^\circ$, $44.6^\circ$, and $75^\circ$ (top to bottom panels), respectively, derived from Eq. (\ref{eq:6}). The vertical dashed lines in the bottom panel denote bifurcation points $B_1$ and $B_2$: at $B_1$, new equilibria $P_4$ and $P_5$ emerge, while at $B_2$, equilibria $P_3$ and $P_4$ merge and annihilate. 
\label{fig:EquSols}}
\end{figure}

 By using Eq. (\ref{eq:6}), we derive the coplanar equilibrium solutions under different primordial obliquities $\psi_{\star0}$, as shown in Fig. \ref{fig:EquSols}. If $\psi_{\star0}$ is low (e.g., $\psi_{\star0}=35^\circ$, top panel), two equilibria two equilibria consistently exist within $(0, 180^\circ)$, denoted as $P_2$ (with $\psi_{1\star}>90^\circ$) and $P_3$ (with $\psi_{1\star}<90^\circ$), respectively. In contrast, if $\psi_{\star0}$ is high (e.g., $\psi_{\star0}=75^\circ$, bottom panel), two bifurcations $B_1$ and $B_2$ emerge within $(0,\psi_{\star 0})$ when $\epsilon_{\star \rm{d}}\sim 1$. At $B_1$, two new equilibria $P_4$ and $P_5$ appear, while at $B_2$, equilibria $P_3$ and $P_4$ merge and annihilate. The critical case $\psi_{\star0}=44.6^\circ$ (middle panel) corresponds to the threshold at which bifurcation begins to occur, as we will demonstrate in the following.

 The specific bifurcation condition can be determined by differentiating eq. (\ref{eq:6}) and setting ${{\mathrm{d}\psi _{1\star}}/{\mathrm{d}\epsilon _{\star \mathrm{d}}=\infty}}$, as illustrated in the bottom panel of Fig. \ref{fig:EquSols}. Equivalently, the bifurcation condition is given by
 \begin{equation}
 	\left\{ \begin{array}{c}
	f\left( \psi _{1\star}; \psi _{\star 0}, \epsilon _{\star \mathrm{d}} \right) =0\\
	\mathrm{d}f\left( \psi _{1\star}; \psi _{\star 0}, \epsilon _{\star \mathrm{d}} \right) / \mathrm{d} \psi_{1\star} =0.\\
\end{array} \right. 
\label{eq:7}
\end{equation}

\begin{figure}
\plotone{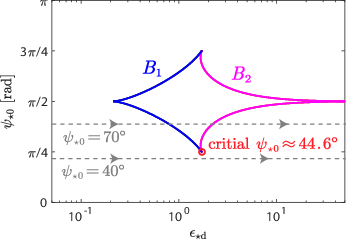}
\caption{Bifurcation curves presented in the ($\psi_{\star0}, \epsilon_{\star \rm{d}}$) parameter space by using the bifurcation condition Eqs. (\ref{eq:7}). The curves for $B_1$ and $B_2$ are distinguished by colors. The critical primordial obliquity for triggering the bifurcation is indicated.
\label{fig:BfCurve}}
\end{figure}

 By using this condition, we obtain the bifurcation curves in the $(\psi_{\star 0}, \epsilon_{\star \rm{d}})$ space, as shown in Fig. \ref{fig:BfCurve}, where we set the disk softening scale $\beta=0.01$. The bifurcation curves are symmetric about $\psi_{\star0}=90^\circ$. These curves delineate the conditions under which bifurcations occur.

 In practical planetary systems, planets are initially dominated by the disk potential, having $\epsilon_{\star \rm{d}}\ll1$. As the the disk dissipates, $\epsilon_{\star \rm{d}}$ increases, eventually reaching $\epsilon_{\star \rm{d}} \gg 1$ once the disk has sufficiently dissipated. Consequently, the occurrence of bifurcations depends solely on the primordial obliquity $\psi_{\star0}$, since $\epsilon_{\star \rm{d}}$ inherently evolves from $\epsilon_{\star \rm{d}}\ll1$ to $\epsilon_{\star \rm{d}}\gg 1$ during disk dissipation. By solving Eqs. (\ref{eq:7}), we find the critical $\psi_{\star0}$ is about $44.6^\circ$. That is, if $\psi_{\star0}<44.6^\circ$, no bifurcation occurs during disk dissipation, while if $\psi_{\star0}<44.6^\circ$, bifurcations will be triggered as $\epsilon_{\star \rm{d}}$ evolves, as the examples of $\psi_{\star0}=40^\circ$ and $70^\circ$ shown in the figure.

\begin{figure}
 \plotone{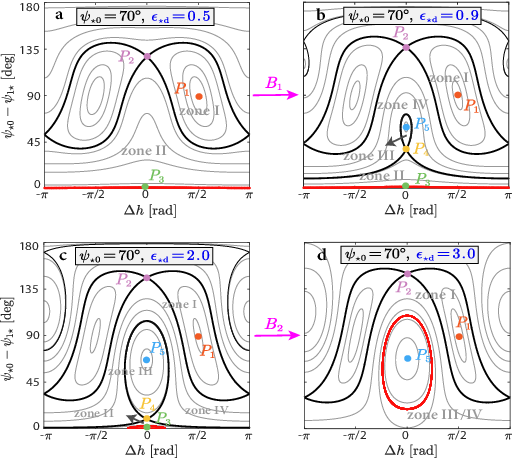}
 \caption{Phase portraits for $\psi_{\star0}=70^\circ$ at $\epsilon_{\star \rm{d}}=0.5$ (a), 0.9 (b), 2.0 (c), and 3.0 (d), respectively. Portraits are plotted in the $(\Delta h, \psi_{\star0}-\psi_{1\star})$ plane, where $\Delta h$ denotes the difference between the longitudes of the planet's orbit and the disk, and $\psi_{\star0}-\psi_{1\star}$ denotes the angle between $\boldsymbol{l}_1$ and $\hat{\boldsymbol{l}}_{\rm{d}}$. In each panel, equilibria are marked by colored points. Note due to the symmetry, if $(\Delta h, \psi_{\star0}-\psi_{1\star})$ is an equilibrium, $(\Delta h-\pi,\ \pi-\psi_{\star0}+\psi_{1\star})$ is also an equilibrium. Trajectories passing through saddle points (bold curves) partition the phase space into distinct libration zones. Red curves show the numerical solution for a fiducial one-planet system at the specified $\epsilon_{\star \rm{d}}$, by using the secular model based on the Laplace-Lagrange theory (Appendix \ref{app_EqOM}). The fiducial system parameters are: $m_{\star}=1M_{\odot}$, $R_{\star}=1.5R_{\odot}$, $P_{\star}=5$ days, $m_1=5m_{\oplus}$, $a_1=0.05$ au, $m_{\mathrm{dk}0}=0.02M_{\odot}$, $r_{\mathrm{in}}=0.03$ au, $r_{\mathrm{c}}=2$ au, $r_{\mathrm{out}}=50$ au, $\tau _{\mathrm{v},\mathrm{in}}=1\times 10^4$ yr, and $\psi _{\star0}=70^\circ$. (Definitions of these parameters can be found in Appendix \ref{app_EqOM})
 \label{fig:PhaPor}}
 \end{figure}

 Due to the integrability of this simplified system, the phase space structure can be fully characterized through phase portraits. Figure \ref{fig:PhaPor} shows phase portraits for a system with high primordial obliquity ($\psi_{\star0}=70^\circ$) at different values of $\epsilon_{\star \rm{d}}$, constructed by plotting energy curves of the secular potential, Eq. (\ref{eq:1}). When $\epsilon_{\star \rm{d}}=0.5$ (panel a), the system exhibits one orthogonal equilibrium $P_1$ and two coplanar equilibria, $P_2$ and $P_3$. The trajectories passing through the saddle point $P_2$ separate two distinct libration regions: Zone I, centered at the orthogonal equilibrium $P_1$, and Zone II, centered at the coplanar equilibrium $P_3$. When $\epsilon_{\star \rm{d}}$ increases to 0.9 (panel b), the system had experienced bifurcation $B_1$, producing new coplanar equilibria $P_4$ and $P_5$. The trajectories through the saddle point $P_4$ separate Zone II around $P_3$, Zone IV around $P_5$, and Zone III encompassing both $P_3$ and $P_5$. As $\epsilon_{\star \rm{d}}$ continues to increase (panel c), the equilibrium $P_4$ migrates toward $P_3$. Finally before $\epsilon_{\star \rm{d}}=3.0$ (panel d), bifurcation $B_2$ triggers the merger and annihilation of $P_3$ and $P_4$, erasing Zone II entirely.

 To elucidate how these bifurcations influence the planet's orbital evolution, we consider the evolution of a fiducial one-planet system. The red curves in each panel of Fig. \ref{fig:PhaPor} depict numerical solutions for this system at the specified values of $\epsilon_{\star \rm{d}}$. Initially, the planet lies within the disk plane, aligning with the equilibrium $P_3$. As the disk disperses, the planet remains in Zone II, librating about $P_3$ with small amplitudes, until encountering bifurcation $B_2$. This bifurcation triggers an abrupt, non-adiabatic transition from Zone II to Zone III, forcing the planet into large-amplitude libration around $P_5$. We refer to this dynamical phenomenon as the dynamical bifurcation-induced effect, highlighting its role in reshaping orbital configurations during disk dissipation.

 \section{Bifurcated Evolutionary Pathways} \label{sec:BEP}
 In the following sections, we explore how the fundamental mechanism--the dynamical bifurcation-induced effect proposed above--shapes the dynamical evolution of multi-planet systems embedded in misaligned protoplanetary disks. Considering an $n$-planet system, assuming all orbits are initially circular, the equations of motion are given by (Appendix \ref{app_EqOM})
 \begin{equation}
 \begin{split}
 	\dot{\boldsymbol{l}}_i&=-\left( \boldsymbol{l}_i\cdot \hat{\boldsymbol{s}}_{\star} \right) \hat{\boldsymbol{s}}_{\star}\times \boldsymbol{l}_i-\sum_{j\ne i}{\frac{1}{\epsilon _{\star j,i}}\left( \boldsymbol{l}_i\cdot \hat{\boldsymbol{l}}_j \right) \hat{\boldsymbol{l}}_j\times \boldsymbol{l}_i}
	\\
	&-\frac{1}{\epsilon _{\star \mathrm{d},i}}\frac{1}{S_i}\left( 1-\frac{\pi}{4}S_i \right) \left( \boldsymbol{l}_i\cdot \hat{\boldsymbol{l}}_{\mathrm{d}} \right) \hat{\boldsymbol{l}}_{\mathrm{d}}\times \boldsymbol{l}_i \,.
 \end{split}
 \label{eq:8}
 \end{equation}
Here $i= \left \{ 1,2,...,n \right \}$;
\begin{equation}
 \epsilon _{\star j,i}=\frac{\phi _{\star, i}}{\phi _{ij}}=\begin{cases}
	\frac{6m_{\star}J_2R_{\star}^{2}a_{j}^{2}}{m_ja_{i}^{4}}\frac{1}{b_{{{3}/{2}}}^{\left( 1 \right)}\left( {{a_i}/{a_j}} \right)},\  a_i<a_j\\
	\frac{6m_{\star}J_2R_{\star}^{2}}{m_ja_ia_j}\frac{1}{b_{{{3}/{2}}}^{\left( 1 \right)}\left( {{a_j}/{a_i}} \right)},\ \ \ a_i>a_j \,, \\
\end{cases}
\label{eq:9}{}
\end{equation}
where the planetary interactions $\phi_{ij}$ is described through the Laplace-Lagrange theory and $b^{(1)}_{2/3}$ is the Laplace coefficient, given by Eq. (\ref{eq:A19}); and 
\begin{equation}
	\epsilon _{\star \mathrm{d},i}=\frac{\phi _{\star ,i}}{\phi _{\mathrm{dk},i}}=\frac{3}{2}\frac{m_{\star}}{2\pi \Sigma _{\mathrm{a}_i}a_{i}^{2}}\frac{J_2R_{\star}^{2}}{a_{i}^{2}} \,.
\label{eq:10}
\end{equation}

Equilibrium solutions are derived by imposing $\dot{\boldsymbol{l}}_i=0$ for all $i={1,2,...,n}$. That is, $3\times n$ coupled nonlinear equations should be solved to obtain the equilibria. While solutions may exist for various dynamical configurations, we restrict our analysis to equilibria that directly govern the planets' evolution. Given the conditions that all planets initially lie within the disk plane, we focus exclusively on coplanar equilibria, parameterized by each planet's stellar obliquity $\psi_{i\star}$. Under these conditions, the equilibria equations simplify to $n$ coupled equations, given by
\begin{equation}
\begin{split}
    &\sin 2\psi _{i\star}-\frac{1}{\epsilon _{\star \mathrm{d},i}}\frac{1}{S_i}\left( 1-\frac{\pi}{4}S_i \right)\sin 2\left( \psi _{\star 0}-\psi _{i\star} \right)
    \\
    &~~-\sum_{j\ne i}{\frac{1}{\epsilon _{\star j,i}}\sin 2\left( \psi _{j\star}-\psi _{i\star} \right)}=0 \,.
\end{split}
\label{eq:11}
\end{equation}
for $i=1,2,..., n$.

\begin{figure*}[ht!]
\plotone{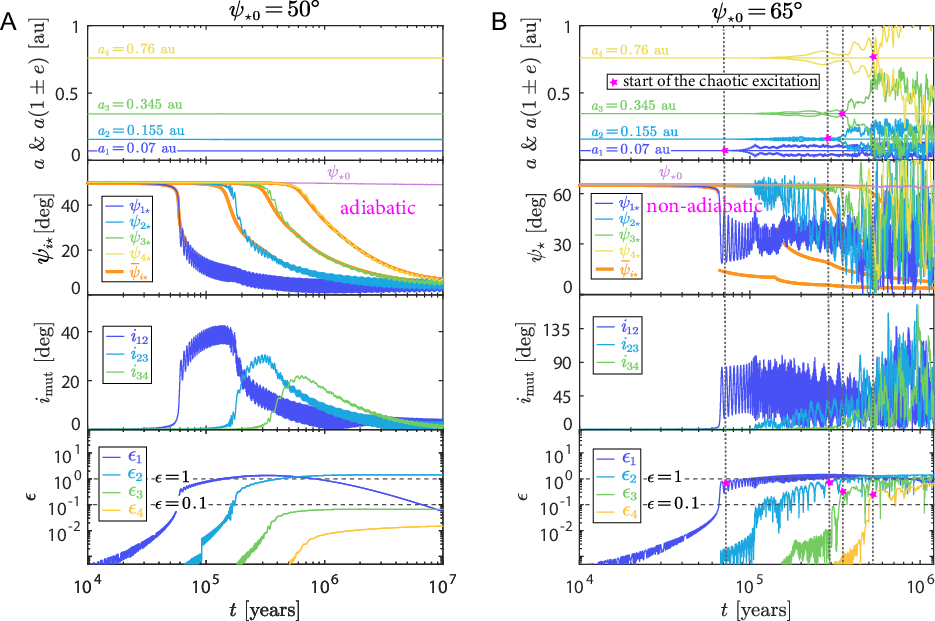}
\caption{Representative dynamical evolution of a compact four-planet system. This figure compares two scenarios featured by different primordial misalignment angles: (A) $\psi _{\star 0}=50^\circ$ and (B) $\psi _{\star 0}=65^\circ$, with identical initial conditions otherwise. Simulations employ the secular model (Appendix \ref{app_EqOM}), which describes the planetary interactions through Gaussian Ring method. In each scenario, panels from top to bottom depict the evolution of the semi-major axes (with pericenter and apocenter distances), stellar obliquities, planetary mutual inclinations, and the frequency ratio of each planet ($\epsilon \sim {{\omega _{\mathrm{in}}}/{\omega _{\mathrm{out}}}}$), respectively. The orange curves in the second panels depict the equilibrium trajectories of the four planets by using Eqs. (\ref{eq:11}). Individual planets are color-coded. The simulations parameters are: $m_{\star}=1m_{\odot}$, $R_{\star}=2R_{\odot}$, $P_{\star}=5$, $k_{\mathrm{q}\star}=0.1$, $m_1\text{-}m_4=5m_{\oplus}$, $m_{\mathrm{d}0}=0.05m_{\star}$, $r_{\mathrm{in}}=0.04$, $r_{\mathrm{out}}=50$, and $\beta=0.01$. The semi-major axes of planetary orbits are indicated in the top panels. The inner disk disperses over a timescale of $\tau _{\mathrm{v}.\mathrm{in}}=10^4$ years, and the stellar oblateness decays over $\tau _{\mathrm{R}_{\star}}=1$ Myr.
\label{fig:FidExa}}
\end{figure*}

By using this condition, the coplanar equilibrium solutions of a four-planet system, considering the disk dissipation and stellar oblateness decay, are depicted by orange curves in the second panels of Fig. \ref{fig:FidExa}(A) and (B) for $\psi_{\star0}=50^\circ$ and $\psi_{\star0}=65^\circ$, respectively. The discontinuous jump in the equilibrium paths for $\psi_{\star0}=65^\circ$ is induced by the bifurcation $B_2$ identified above. Note that, while $\psi_{\star0}=50^\circ$ exceeds the critical obliquity ($\psi_{\star0}\approx44.6^\circ$) established for one-planet system, no bifurcation occurs in this multi-planet configuration. This demonstrates that the interplanetary gravitational couplings increase the critical obliquity in multi-planet systems. A quantitative analysis of these couplings and their dynamical consequences lies beyond the scope of this work and will be focused in our future work.

Figure \ref{fig:FidExa} also shows the complete dynamical evolution of the four-planet system by using the full secular model detailed in Appendix \ref{app_EqOM}, where the planetary interactions are described through the Gaussian Ring algorithm. As illustrated, in the scenario of $\psi_{\star0}=50^\circ$, all the planets adiabatically track their respective equilibria with no eccentricity excitation. As a result, with the dissipation of the disk, planets smoothly transition from their initial alignment with the disk to near alignment with the stellar spin axis, ultimately producing a final architecture characterized by near-zero mutual inclinations ($i_{\mathrm{mut}}\sim 0$) and minimal stellar obliquity ($\psi _{i\star}\sim 0$).

In contrast, when $\psi_{\star0}=65^\circ$, an abrupt, non-adiabatic transition is triggered due to the dynamical bifurcation-induced effect, initiated at the inner-most orbit. This excites large-amplitude librating mutual planetary inclinations, which subsequently trigger eccentricity excitation in the inner orbit. Crucially, this process can propagate from inner to outer orbits during the disk dissipation, forming an excitation chain, and finally destabilizes the entire system.

\begin{figure*}[ht!]
\plotone{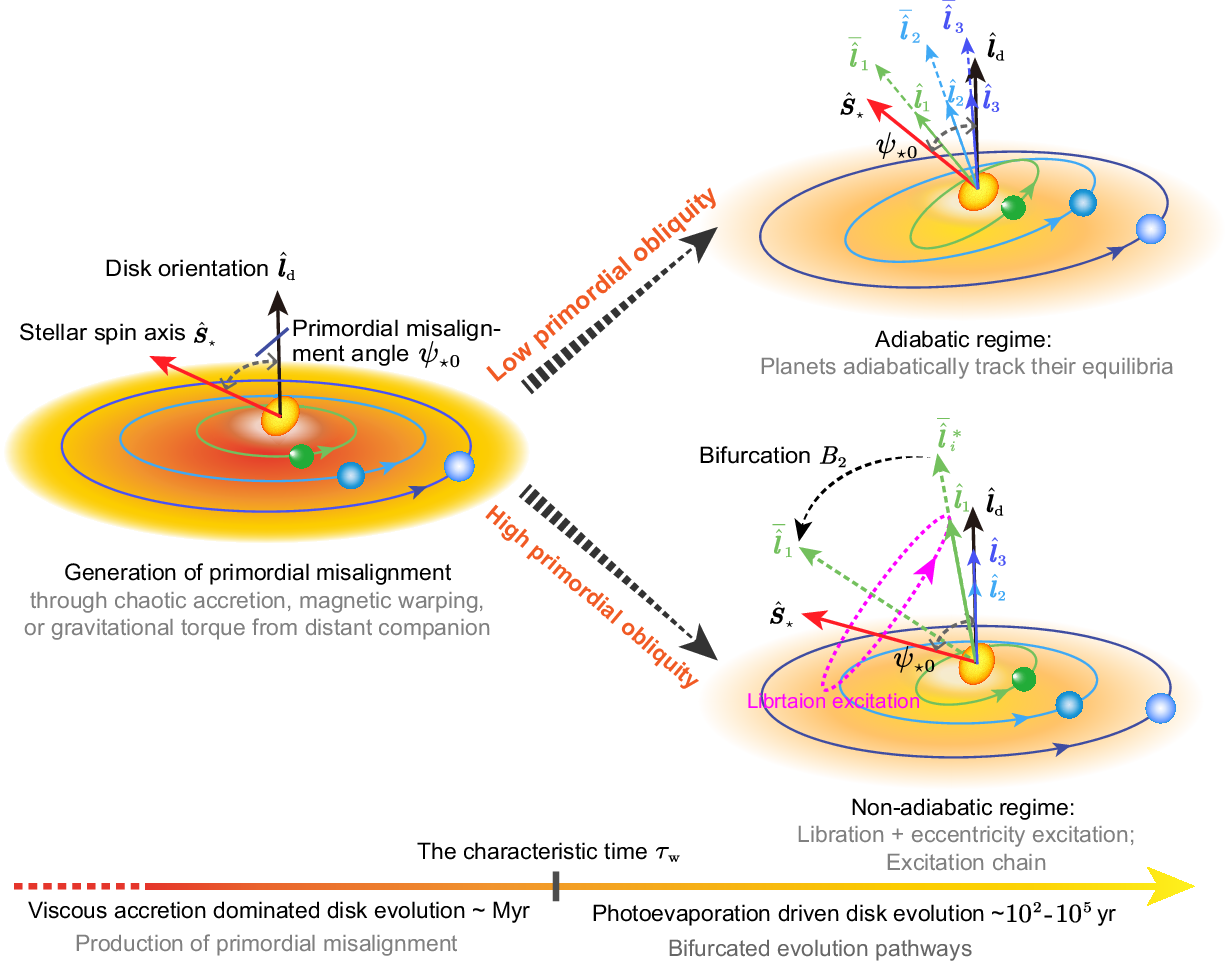}
\caption{Schematic of the bifurcated evolution pathways driven by misaligned protoplanetary disks. Before the characteristic time $\tau_{\rm{w}}$, the misalignment between the disk $\hat{\boldsymbol{l}}_{\mathrm{d}}$ and the stellar spin axis $\hat{\boldsymbol{s}}_{\star}$ is produced, with the misalignment angle $\psi _{\star 0}$. After $\tau_{\rm{w}}$, photoevaporation drives the rapid dissipation of the disk over a timescale of $10^2\text{-}10^5$ yr. During this stage, bifurcated evolutionary paths occur based on the extent of the primordial misalignment: planets in systems with low $\psi_{\star0}$ adiabatically follow their evolving equilibria ($\bar{\hat{\boldsymbol{l}}}_i$), smoothly transitioning the alignment from the disk to the stellar spin axis (upper right); conversely, systems with high $\psi_{\star0}$ encounter an instant, adiabatic transition ($\bar{\hat{\boldsymbol{l}}}_{i}^{*}\rightarrow \bar{\hat{\boldsymbol{l}}}_i$) due to the dynamical bifurcation-induced effect, leading to large-amplitude librating angle, as illustrated in the lower right plot. The significant libration excitation triggers the excitation of eccentricity, and such process can propagate from inner to outer orbits, forming a excitation chain, and finally destabilizes the entire system.
\label{fig:Geostru}}
\end{figure*}

Consequently, the two dynamical regimes--adiabatic (e.g., $\psi_{\star0}=50^\circ$) and non-adiabatic (e.g., $\psi_{\star0}=65^\circ$)--primarily dependent on the misalignment angle $\psi_{\star0}$, yield bifurcated evolutionary pathways in multi-planet systems and thereby distinct final orbital architectures. Figure. \ref{fig:Geostru} provides a schematic illustration of our proposed framework. In the adiabatic regime, systems with low primordial stellar obliquities undergo mild evolution, forming final coplanar and well-aligned configurations. Conversely, in the non-adiabatic regime, the bifurcation-induced effect excites large-amplitude librations, which drive eccentricity excitation. This instability propagates outward from the inner to outer orbits, generating an excitation chain that ultimately destabilizes the system.

\section{Excitation Chain in the Non-Adiabatic Regime} \label{sec:CEC}
 As studied above, in the adiabatic regime, planetary systems behave in a mild and predictable way, whereas in the non-adiabatic regime, the dynamics becomes dramatic and complex. We have already demonstrated that the dynamical bifurcation-induced effect is responsible for initiating the non-trivial dynamical process in the non-adiabatic regime. However, what mechanism drives eccentricity excitation, and how the excitation propagate across orbits to form an excitation chain, remain unsolved. This section elucidates the physical processes underlying these phenomena.

 First, the eccentricity excitation is studied by numerically exploring the parameter space defined by the dimensionless parameter $\epsilon ={{\omega _{\mathrm{out}}}/{\omega _{\mathrm{in}}}}$ and the stellar obliquity $\psi_{\star0}$. Here, $\omega_{\rm{in}}$ represents the in-plane orbital precession frequency, driven by perturbations such as gravitational interactions with the protoplanetary disk or co-planar planets, and $\omega_{\rm{out}}$ corresponds to the out-of-plane precession frequency, induced by the gravitational influence of the inclined stellar oblateness or planets perturbed out of the disk plane. For single-planet systems, $\epsilon$ reduces to $\epsilon_{\star\rm{d}}$ defined by eq. (\ref{eq:5}), where the dynamical-bifurcation effect requires $\epsilon_{\star \rm{d}} \gtrsim 1$. Analogously, the dynamical bifurcation also requires $\epsilon$ to reach a value of $\gtrsim 1$.

 The necessary condition for the eccentricity excitation is the large libration about equilibrium induced by the dynamical bifurcation-induced effect, at which the disk's gravity becomes relatively small compared to the stellar oblateness (i.e., $\epsilon_{\star \rm{d}}\gg1$, see bottom panel of Fig. \ref{fig:EquSols}). In addition, with the subsequent rapid dissipation of the disk, the disk's gravitational influence becomes negligible compared to the stellar oblateness. We therefore neglect disk gravity in our analysis of the eccentricity excitation mechanism for simplicity.

\begin{figure} [ht!]
 \plotone{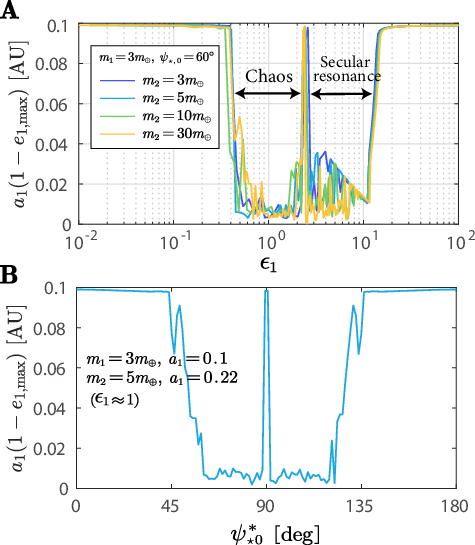}
 \caption{ Eccentricity excitation as functions of $\epsilon$ (panel A) and $\psi^{*}_{\star0}$ (panel B), illustrated for two-planet systems. Simulations employ the secular model based on the Gaussian Ring method, with planetary orbits initially coplanar with a given initial obliquity $\psi^{*}_{\star0}$. The stellar parameters are $m_{\star}=m_{\odot}$, $R_{\star}=1.5R_{\odot}$, and $P_{\star}=3$ days, which yields $J_2\approx 3.4\times 10^{-4}$. Panel A shows eccentricity excitation versus $\epsilon_1$. In this calculation, four cases of planet mass ratios are considered, with $\psi _{\star 0}=60^\circ$, $m_1=3m_{\oplus}$, and $a_1=0.1$ au held constant. By systematically varying the semi-major axis $a_2$, $\epsilon_1$ ranges from $10^{-2}$ to $10^{2}$. The maximum eccentricity, ${e_{1,}}_{\max}$, is recorded during the integrations up to 0.5 Myr for each integration case of $\psi _{\star 0}$. Panel B shows eccentricity excitation versus $\psi^{*}_{\star0}$ at fixed $\epsilon_1=1$. Note: the narrow interval near $\psi^{*}_{\star 0}=90^\circ$ arises from weakened stellar oblateness effects due to the $\left( \hat{\boldsymbol{s}}_{\star}\cdot \hat{\boldsymbol{l}} \right) $ term in the secular equations.
 \label{fig:EccExc}}
 \end{figure}

 Figure \ref{fig:EccExc} maps the parameter space for eccentricity excitation as functions of both $\epsilon$ and $\psi^{*}_{\star0}$. Here, to distinguish it from the primordial star-disk misalignment angle $\psi_{\star0}$ (considering disk), we denote the star-planet misalignment angle as $\psi^{*}_{\star0}$ (without disk). Panel A (where $\psi^{*}_{\star0}=60^\circ$) shows that the eccentricity excitation is triggered at $\epsilon \sim 1$, where the stellar oblateness is comparable to the planetary interactions. Intriguingly, we also identify another excitation region featured by $\epsilon \gg 1$, where the stellar oblateness dominates over the planetary interactions.

 Our analysis of Poincaré surfaces of section (Fig. \ref{fig:PoinSec}, Appendix \ref{app_EccExc}) reveals that the two excitation regimes originate from fundamentally different mechanisms: the $\epsilon \sim 1$ excitation stems from dynamical chaos (overlapping secular resonances), which operates for all initial obliquities $\psi^{*}_{\star0} \gtrsim 45^\circ$ (see panel B of Fig. \ref{fig:EccExc}); and the $\epsilon \gg 1$ excitation arises from a particular class of secular resonance that requires specific high initial obliquities \citep{yokoyama1999dynamics}. For our purpose, it is sufficient to focus on the excitation driven by the chaos at $\epsilon \sim 1$. We refer to this dynamical effect as the chaotic eccentricity excitation mechanism.

 Panel B reveals that the critical obliquity for the chaotic eccentricity excitation is $\psi^{*}_{\star0}\approx45^\circ$. In the absence of disk gravity, this obliquity ensures sufficiently large libration amplitudes about equilibrium to enable eccentricity excitation. On the other hand, according to the results in Sec. \ref{sec:BEP}, such large libration angles in misaligned protoplanetary disk systems can only be achieved in the non-adiabatic regime via dynamical bifurcation-induced effect. Recall that the critical primordial obliquity for this effect is $\psi_{\star0} \approx 44.6^\circ$ for single-planet system, with marginally larger values for multi-planet systems. This close agreement suggests that in most non-adiabatic evolutions, the bifurcation-induced libration amplitude will be adequate to trigger the chaotic eccentricity excitation.

  We further validate our findings by examining three-planet systems in Fig. \ref{fig:EccExc2} (Appendix \ref{app_EccExc}). The robust consistency of these results across different system architectures and parameter combinations supports the universality of the chaotic excitation mechanism we have identified.

  This libration excitation and then eccentricity excitation process can propagate outward through the system, forming an excitation chain. This excitation chain is established as follows. First, in systems with high $\psi_{\star0}$, for the inner-most planet, $\omega_{\rm{in}}$ is contributed by the disk's gravity and perturbations from the outer planets, while $\omega_{\rm{out}}$ arises from the inclined stellar oblateness. Initially, the ratio $\epsilon_1={{\omega _{\mathrm{out}}}/{\omega _{\mathrm{in}}}} \ll 1$. As the disk disperses, this ratio increases until reaching a critical threshold ($\epsilon_1 \gtrsim 1$), where the dynamical bifurcation-induced effect drives the innermost planet into a high-inclination, rapidly librating state. The resulting inclined orbit of the inner planet then enhances $\omega_{\rm{out}}$ for adjacent outer planets, effectively acting as an additional out-of-plane perturber. When the frequency ratio  ($\epsilon_2={{\omega _{\mathrm{out}}}/{\omega _{\mathrm{in}}}}$) for the second inner orbit similarly reaches the critical value, the same  excitation process is activated. Following this manner, this process consequently propagates outward through the system, forming an excitation chain. Figure \ref{fig:FidExa}B illustrates such an example. Note the extreme eccentricity growth of inner orbits introduces significant chaotic perturbations to outer orbits, which may obscure certain details of the excitation chain.

 \section{Statistical Analysis} \label{sec:SA}

 In this section, we evaluate the prevalence of the proposed framework (Fig. \ref{fig:Geostru}) through a statistical analysis of the \textit{Kepler} sample of compact multi-planet systems. Recall that systems with low $\psi_{\star0}$ generally undergo trivial adiabatic evolution, while systems with high $\psi_{\star0}$ could experience non-adiabatic evolution, involving complex dynamical processes--particularly the excitation chain. Consequently, our analysis focuses primarily on the prevalence of the non-adiabatic regime, which primarily depends on two factors: (1) the occurrence rate of highly misaligned protoplanetary disks, and (2) the fulfillment of the conditions required for the excitation chain. Current evidence suggests that a significant fraction of protoplanetary disks can be moderately misaligned (see Sec. \ref{sec:intro}). Here, we investigate whether the excitation chain can be established in typical compact multi-planet systems.

 Establishing an excitation chain within a compact multi-planet system requires two key steps. First, the excitation must be initiated in the inner-most orbit, which corresponds to triggering the dynamical bifurcation-induced effect. This necessitates a maximum ratio of $\epsilon_{1,\rm{max}} \gtrsim 1$, which is achieved assuming the disk has sufficiently dissipated. Second, the excitation must propagate outward, requiring that the maximum $\epsilon_{i,\rm{max}} \gtrsim 1$ for outer planets ($i>1$).
 
\begin{figure*}[htb!]
 \plotone{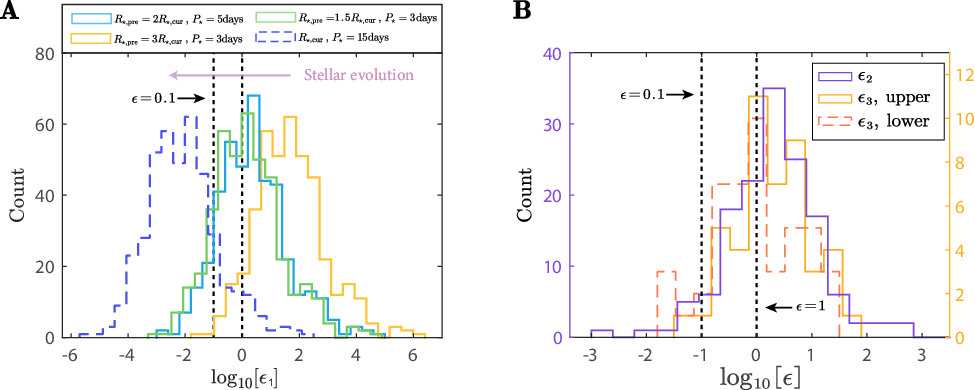}
 \caption{Statistics of the frequency ratio ($\epsilon$) for \textit{Kepler} compact multi-planet systems. Our sample includes 1052 planets in 414 systems (with $n \geq 2$) from the \textit{Kepler} DR25 catalog, with stellar parameters and planetary masses retrieved from \href{https://exoplanetarchive.ipac.caltech.edu/} {NASA Exoplanet Archive}. Panel A shows statistics for $\epsilon_{1,\rm{max}}$. The blue dashed histogram presents the distribution using current stellar radii, $R_{\star ,\mathrm{cur}}$, and assuming $P_{\star}=15$ days. The other three histograms represent systems with three different typical pre-main-sequence stellar parameters: a strong stellar oblateness case (yellow) and two weaker oblateness cases ( green and sky-blue), as specified in the figure. Panel B shows statistics for $\epsilon_{2,\rm{max}}$ ($n \geq 3$ systems, left y-axis) and $\epsilon_{3,\rm{max}}$ ($n \geq 4$ systems, right y-axis), calculated by assuming $R_{\star}=2R_{\star ,\mathrm{cur}}$ and $P_{\star}=3$ days. For $\epsilon _3$, both the lower ($\omega_{\rm{out}}$ from adjacent inner orbit only) and upper ($\omega_{\rm{out}}$ from all inner orbits and stellar oblateness) limits are depicted.
 \label{fig:Statis}}
 \end{figure*}

 Figure \ref{fig:Statis} presents the statistics over the \textit{Kepler} sample of multi-planet systems from the final \textit{Kepler} data release (DR25) \footnote{Data retrieved from the NASA Exoplanet Archive (\href{https://exoplanetarchive.ipac.caltech.edu}{exoplanetarchive.ipac.caltech.edu})} \citep{thompson2018planetary}. Panel A shows the distribution of $\epsilon_{1,\rm{max}}$, where we adopt several sets of typical pre-main-sequence stellar parameters, based on the observations that T Tauri stars typically possess short rotation periods ranging from 1 to 10 days and distended radii \citep{bouvier2014angular}. As observed, the current distribution of $\epsilon_{1,\rm{max}}$ falls predominantly below the critical threshold $\epsilon_1 \sim 1$. However, during the pre-main-sequence stage, a substantial portion of systems exhibit $\epsilon_1 \gtrsim 1$. Specifically, approximately $90\%$ of the systems have $\epsilon _1\gtrsim 1$ in the strong oblateness case, while a similar fraction in the two weaker cases yields $\epsilon _1>0.1$, which is close to the lower limit of the excitation.
 
 Panel B shows distributions of the frequency ratios for the second and third inner orbits ($\epsilon_2$ and $\epsilon_3$), which reflect the efficiency of the propagation of the excitation. The distributions of $\epsilon_2$ and $\epsilon_3$ peak at $\epsilon \sim 1$, suggesting that once initiated, the excitation chain tends to be robust. This feature is fundamentally linked to the “peas-in-a-pod” characteristics of the \textit{Kepler} multi-planet systems \citep{millholland2017kepler, weiss2018california, he2019architectures, goyal2024peas}, which facilitates the frequency commensurability condition ($\epsilon \sim 1$) within the outer orbits.

 Consequently, we conclude that if primordial misalignment between protoplanetary disks and stellar spin axes is common, the dynamical framework illustrated in Fig. \ref{fig:Geostru} likely governs the evolution of most compact multi-planet systems. That is, systems with high $\psi_{\star0}$ typically undergo dramatic, non-adiabatic evolution through the excitation chain, leading to strong instability and system destruction. As a result, only compact multi-planet systems with low $\psi_{\star0}$, experiencing mild and adiabatic evolution, can survive. This framework naturally predicts that compact multi-planet systems are finally characterized by coplanar and low stellar obliquity, while coplanar multiple planets with high stellar obliquity would be rare. These predictions align well with current observations that \textit{Kepler} multi-planet systems generally have low stellar obliquities \citep{winn2017constraints, munoz2018statistical}.
 
 On the other hand, systems featuring a maximum frequency ratio of $\epsilon _1\ll 1$ are more likely to avoid the excitation chain and thus survive from highly misaligned disks. There are two typical scenarios that often meet this condition: one involves a rather small host star, characterized by a weak stellar quadrupole moment; the other involves quite strong gravitational couplings among planets, either due to extremely compact planetary configurations or the presence of a massive close-in planet that enhances such couplings. Therefore, we expect the coplanar multi-planet systems observed with high obliquities are typically associated with either a small host star, or a tightly compact configuration, or a close-in giant planet. However, some other factors may complicate these scenarios. For example, a weak stellar quadrupole moment may be not favorable for the formation of high primordial obliquity \citep{zanazzi2018planet} and extremely packed planets are prone to long-term instability \citep{chambers1996stability,zhou2007post,quillen2011three,pu2015spacing}. In the following, we present two examples, K2-290 A and HD 3167, that probably survived from their highly misaligned protoplanetary disks.

 \section{Sculpting of Compact Multi-Planet Systems} \label{sec:SCMPS}

 In this section, we show two examples of non-typical multi-planet systems that potentially shaped by the dynamical framework proposed in Fig. \ref{fig:Geostru}, and then we perform a numerical investigation on the distribution of final stellar obliquity and mutual planetary inclinations sculpted by this framework.

 \subsection{K2-290A and HD 3167}
 
 The observed high stellar obliquities in the K2-290 A and HD 3167 systems have been previously attributed to primordial star-disk misalignments \citep{hjorth2021backward, fu2024sculpting}. Therefore, these systems could serve as ideal candidates for examining the dynamical process proposed in this study.

 The K2-290A system hosts two coplanar planets in a retrograde configuration relative to the stellar spin, with an obliquity of $\psi _{\star}\approx 124^\circ\pm 6^\circ$. This architecture has been recently interpreted as evidence for primordial misalignment, where the stellar companion K2-290B is responsible for having tilted the protoplanetary disk to $\psi _{\star 0}\approx \psi _{\star}$ \citep{hjorth2021backward}. 

 However, the previous study focus solely on the disk-tilting phase (before $\tau_{\rm{w}}$ in Fig. \ref{fig:Geostru}), neglecting the subsequent evolution during rapid photoevaporation-driven disk dispersal (after $\tau_{\rm{w}}$). Theoretically, maintaining coplanarity between these two planets requires adiabatic evolution. But, according to our framework, such a high primordial obliquity ($\psi_{\star0}\approx124^\circ$) likely triggers dramatic, non-adiabatic evolution. If so, the observed stellar obliquity of K2-290 A would be incompatible with the primordial misalignment scenario, and alternative formation channels are required \citep{best2022chaotic}.

\begin{figure}
 \plotone{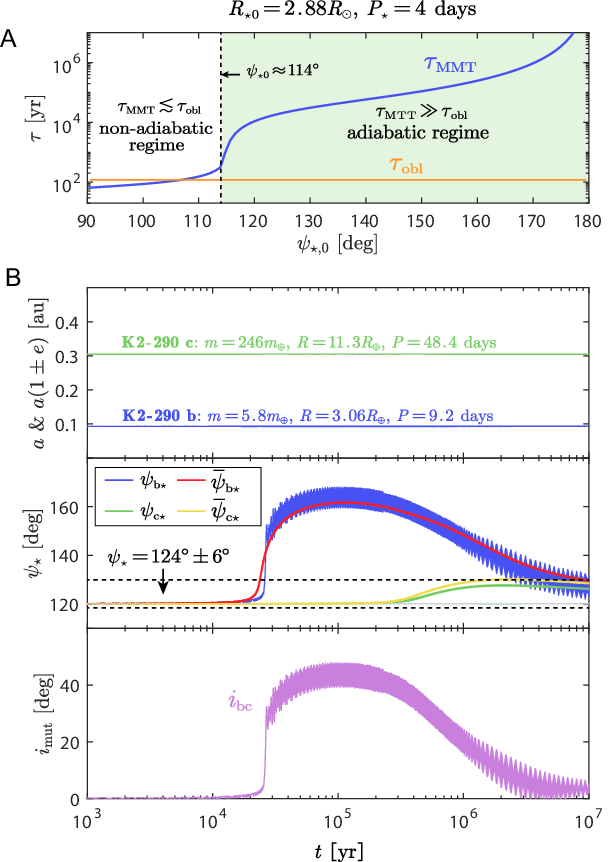}
 \caption{Dynamical evolution of the K2-290A system with a highly misaligned protoplanetary disk. This figure extends the work by \citet{hjorth2021backward}. Panel A shows the minimum transition timescale ($\tau_{\rm{MMT}}$) during the disk dissipation and the timescale of the innermost planet's precession induced by the stellar oblateness ($\tau_{\rm{obl}}$). The pre-main-sequence stellar parameters are set as $R_{\star}=2.88R_{\odot}$ and $P_{\star}=4$ days, yielding a strong stellar oblateness $J_2\approx 1.1\times 10^{-3}$. Panel B shows the dynamical evolution of K2-290A by using the secular model based on the Gaussian Ring method, setting $\psi_{\star0}=120^\circ$. From top to bottom, it shows the semi-major axes (including pericenter and apocenter distances), stellar obliquity (including the evolution of equilibria $\bar{\psi}_{\mathrm{b}\star}$ and $\bar{\psi}_{\mathrm{c}\star}$), and mutual inclination between the two planets "b" and "c". The parameters of K2-290A are adopted from \citet{hjorth2021backward}, and other parameters are $\tau _{\mathrm{R}\star}=1$ Myr, $k_{\mathrm{q}\star}=0.1$, $k_{\star}=0.2$, $m_{\mathrm{d}0}=0.05m_{\odot}$, $r_{\mathrm{in}}=0.03$, $r_{\mathrm{c}}=1.0$ au, $\tau _{\mathrm{v},\mathrm{in}}=10^4$, and $\tau _{\mathrm{v},\mathrm{out}}=1$ Myr.
 \label{fig:K2-290A}}
 \end{figure}

 Nevertheless, the presence of the outer massive warm Jupiter (planet "c") significantly enhances planet-planet coupling, significantly raising the critical $\psi_{\star0}$ to trigger the bifurcation-induced effect, as shown in Fig. \ref{fig:K2-290A}. Here, we employ a numerical method, by comparing the minimum transition timescale of equilibrium ($\tau_{\rm{MMT}}$) and the orbital precession timescale ($\tau_{\rm{obl}}$), to determine this critical primordial obliquity\footnote{This critical angle can be analytically solved in hierarchical planetary systems, which will be studied in our later work.}. If $\tau_{\rm{MMT}}\gtrsim\tau_{\rm{obl}}$, the evolution is adiabatic, otherwise the evolution is non-adiabatic.

 Through this method, the critical $\psi_{\star0}$ is determined to be about $114^\circ$, even when assuming a relatively strong stellar oblateness (a weaker stellar oblateness leads to a even higher critical $\psi_{\star0}$ closing to $90^\circ$), as shown in Fig. \ref{fig:K2-290A}. This implies that if $\psi_{\star0}>114^\circ$, the evolution is adiabatic. Numerical simulations for $\psi_{\star0}=120^\circ$ is shown in Fig. \ref{fig:K2-290A}B, which confirms our analysis. During the disk's photoevaporation, the planets are able to roughly adiabatically follow their equilibria, because the dynamical bifurcation is avoided. Furthermore, the outer planet's orbital angular momentum dominates over the stellar spin angular momentum, allowing it to effectively preserve the protoplanetary disk's orientation. As a result, considering the decay of the stellar oblateness, the coplanar and retrograde configuration of the two planets in K2-290A can be reproduced, supporting primordial misalignment as a plausible formation scenario.

 In contrast to the K2-290A system, which requires an adiabatic process to maintain coplanarity, the HD 3167 system necessitates the dynamical bifurcation-induced effect to produce its near-perpendicular mutual planetary inclination. While this non-adiabatic evolution is triggered in the innermost planet, general relativistic (GR) effects suppress eccentricity excitation. Moreover, the excitation chain fails to propagate between the innermost and second innermost orbits. Therefore, HD 3167 remain stable during the non-adiabatic evolution. The detailed discussion is presented in the recent work \citep{fu2024sculpting}.

 Consequently, the K2-290 A and HD 3167 systems exemplify the adiabatic and the non-adiabatic evolution, respectively. However, these two systems are atypical among compact multi-planet systems due to their unique properties: K2-290A holds a short-period giant planet, and the planetary orbits in HD 3167 are not regularly spaced, with $H_{12}\approx 55$ Hill radius, while both $H_{23}$ and $H_{34}$ are smaller than 30 Hill radii. Thanks to their unusual characteristics, these systems provide valuable insights into primordial misalignments. In contrast, typical compact multi-planet systems likely erase such traces, as we demonstrate next.

 \subsection{Typical Compact Multi-Planet Systems}
\begin{figure}[ht!]
 \plotone{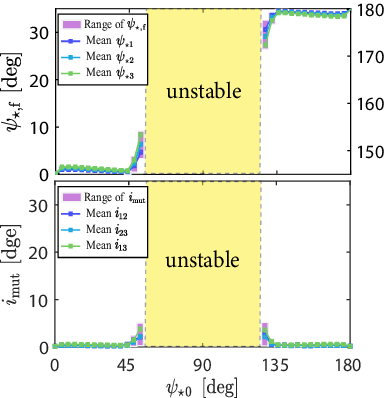}
 \caption{Distributions of final stellar obliquities and mutual inclinations of typical compact multi-planet systems sculpted by misaligned protoplanetary disks, as a function of $\psi_{\star0}$. The simulations are performed over three-planet systems with $m_1=m_2=m_3=5m_{\oplus}$, $a_1=0.05$ au, and the separations between planets are set as 20 Hill radii. The primordial obliquity $\psi_{\star0}$ is uniformly sampled across 46 values within the range over $\left[ 0,180^\circ \right] $. The other parameters are the same to those used in Fig. \ref{fig:FidExa}. Each simulation runs until either 5 Myr is reached or orbit crossings between adjacent planet pairs occur, with the latter scenario labeled as unstable. The bars in the upper panel represent the range of final obliquities, determined by the maximum and minimum obliquities recorded during the last $10^4$ years of the integration. The curves represent the mean obliquity of each planet over this period. The right y-axis applies to the cases of $\psi _{\star0}>90^\circ$. The lower panel displays the distribution of the final mutual inclinations, with similar settings with the upper panel. 
 \label{fig:Distri}}
 \end{figure}

 We perform numerical simulations over a set of three-planet systems with $\psi_{\star0}$ spanning $0$ to $180^\circ$. The resulting final stellar obliquities and mutual inclinations are shown in Fig. \ref{fig:Distri}, where systems undergoing orbit crossings due to the excitation chain are labeled as unstable. For the adopted system parameters, systems with $\psi_{\star0}$ in the range $52^\circ\text{-}128^\circ$ typically experience the excitation chain. Outside this range, systems evolve adiabatically, settling into near-coplanar, low-obliquity configurations, with both obliquities and mutual inclinations deviating by only a few degrees from $0^\circ$ (or $180^\circ$ for obliquities). This distribution is slightly influenced by the timescale of the disk dissipation $\tau_{\rm{v,in}}$. Compared to the fiducial value ($\tau_{\rm{v,in}}=10^4$ yr) used in Fig. \ref{fig:Distri}, shorter timescales produce moderately broader range of final obliquities and mutual inclinations. For instance, when $\tau_{\rm{v,in}}=5\times10^2$ year, our simulations show these parameters can reach up to $30^\circ$. 

 These results show good agreements with current observational constraints on compact multi-planet systems, where coplanar multi-planet systems typically exhibit low stellar obliquities \citep{winn2017constraints, munoz2018statistical}.

 \section{Conclusions} \label{sec:conclusions}
 
 In this study, we have investigated the dynamical evolution of planetary systems driven by misaligned protoplanetary disks. We demonstrate that, based on the dynamical bifurcation-induced effect, multi-planet systems with misaligned protoplanetary disks typically undergo bifurcated evolutionary pathways, depending on the extent of the primordial star-disk misalignment, as illustrated Fig. \ref{fig:Geostru}.

 The dynamical bifurcation-induced effect arises from the combined influence of the full-space disk potential and stellar oblateness during disk dispersal. This effect triggers an abrupt, non-adiabatic transition of the equilibrium direction, leading to large-amplitude libration of orbit angular momentum. While this effect was first identified by \citet{fu2024sculpting} for polar primordial obliquities, our work extends the analysis to the full range of $\psi_{\star0}$. We identify a critical angle for this effect being about $44.6^\circ$ in one-planet system. That is, if $\psi_{\star0}>44.6^\circ$, this effect occurs. This critical angle increases for multi-planet systems.

 Based on this effect, systems with low and high primordial obliquities exhibit distinct (adiabatic or non-adiabatic) evolution behaviors. In the adiabatic regime (low $\psi_{\star0}$), planets adiabatically follow their respective evolving equilibria, forming final coplanar and low-obliquity configuration. In the non-adiabatic regime, the dynamical bifurcation-induced effect causes large-amplitude libration in orbit, which will trigger chaotic eccentricity excitation if the libration angle excess a critical value. This process propagates outward via an excitation chain, potentially destabilizing the entire system.
 
 Statistical analysis of the \textit{Kepler} sample of compact multi-planet systems confirms the prevalence of this framework. Our numerical simulations further show that surviving systems are predominantly coplanar with low stellar obliquities, aligning with observational constraints.

 The K2-290 A and HD 3167 systems, with their unusual architectures and high stellar obliquities, serve as atypical but illustrative examples of the adiabatic and non-adiabatic regimes, respectively. These systems provide critical insights into the role of primordial misalignments in early planetary evolution.

\begin{acknowledgments}
We thank Will M. Farr for contributing to the public Rings code (\url{https://github.com/farr/Rings}). This work was supported by the Fundamental Research Funds for the Central Universities.
\end{acknowledgments}

%




\appendix

\section{Equations of Motion and Definitions} \label{app_EqOM}

Consider a system of $n$ planets with a fast-rotating young host star (orientated along $\hat{\bm{s}}_{\star}$) and a protoplanetary disk (orientated along $\hat{\bm{l}}_{\rm{d}}$). The Hamiltonian function governing the planetary orbital evolution is given by 
\begin{equation}
\mathcal{H} =\sum_{i=1}^n{\mathcal{H} _{\mathrm{K},i}+\Phi _{\star ,i}+ \Phi_{\mathrm{GR},i} + \Phi _{\mathrm{dk},i}}+\sum_{i < j}{\Phi _{ij}}
\label{eq:A1}
\end{equation}
 where the subscript "$i$" ("$j$") denotes the $i$-th ($j$-th) planet from inner to outer, $\mathcal{H} _{\mathrm{K},i}$ represents the Hamiltonian of Kepler motion for planet $i$, $\Phi _{\star ,i}$ represents the potential due to the stellar quadrupole moment, $\Phi _{\mathrm{dk},i}$ denotes the potential due to the disk's gravity, and $\Phi _{ij}$ denotes the potential due to the gravitational interactions between planets $i$ and $j$. (Explicit expressions of these potentials are given in the following).

In order to study the secular evolution of planetary orbits, averaging method is utilized to derive the secular pattern of these potentials \citep{tremaine2014earth,wang2020semi, fu2023semi}. After averaging, the potentials can be formed in terms of the normalized angular momentum vector $\boldsymbol{l}=\sqrt{1-e^2}\hat{\boldsymbol{l}}$ and the eccentricity vector $\boldsymbol{e}=e\hat{\boldsymbol{e}}$.

The secular potential of the stellar oblateness $\Phi _{\star ,i}$ is given by \citep{wang2020semi}
\begin{equation}
\Phi _{\star ,i}=\frac{\phi _{\star ,i}}{6\left( 1-e_{i}^{2} \right) ^{3/2}}\left[ 1-3\left( \hat{\boldsymbol{s}}_{\star}\cdot \hat{\boldsymbol{l}}_i \right) ^2 \right] 
\label{eq:A2}
\end{equation}
where
\begin{equation}
	\phi _{\star ,i}=\frac{3Gm_{\star}m_iJ_2R_{\star}^{2}}{2a_{i}^{3}}
\label{eq:A3}
\end{equation}
where $G$ is the universal gravitational constant, $m_{\star}$ and $R_{\star}$ are stellar mass and radius, $J_2$ is the coefficient of stellar oblateness, and $m_i$ and $a_i$ are the mass and orbital semi-major axis of planet $i$. 

The rotation-induced stellar oblateness $J_2$ can be estimated by
\begin{equation}
J_2=\frac{2k_{\mathrm{q}\star}\Omega _{\star}^{2}R_{\star}^{3}}{3Gm_{\star}}
\label{eq:A4}
\end{equation}
Here, $k_{\mathrm{q},\star}$ is the stellar apsidal motion constant, where $k_{\mathrm{q},\star} \simeq 0.1$ for fully convective stars, and $\Omega_{\star}$ is the stellar rotation angular velocity. Pre-main sequence (PMS) stars generally have distend radii and short rotation periods (typically 1-10 days for T Tauri stars \citep{bouvier2014angular}), and according to Eq. (\ref{eq:A4}), they have strong stellar oblateness, typically resulting in $J_2\sim10^{-2}\text{-}10^{-4}$ \citep{spalding2020stellar}.

During the PMS stage, the stellar oblateness decays as a result of the stellar contraction, spin-down, evolution of $k_{\mathrm{q},\star}$. According to Eqs. (\ref{eq:A3}) and (\ref{eq:A4}), the effect of stellar oblateness is quite sensitive to $R_{\star}$. Therefore, we describe the oblateness decay through including the contraction of $R_{\star}$ in a way (the specific manner of stellar oblateness decay is not important, given the change is adiabatic)
\begin{equation}
R_{\star}\left( t \right) =R_{\star}\left( 0 \right) \left( 1+\frac{t}{\tau _{R_{\star}}} \right) ^{{{-1}/{3}}}
\label{eq:A5}
\end{equation}
where $\tau_{R_\star}=1$ Myr, roughly coinciding with the Kelvin-Helmholtz timescale \citep{batygin2013magnetic}.

The secular potential of general relativity effect, $\Phi_{\mathrm{GR},i}$, is given by
\begin{equation}
\Phi _{\mathrm{GR},i}=-\frac{3G^2m_{\star}^{2}m_i}{a_{i}^{2}c^2\sqrt{(1-e^2_{i})}}
\label{eq:A6}
\end{equation}
where $c$ is the speed light.

For disk, the potential of the Mestel disk's gravity utilized in this study is derived form that of the homeoidal disk \citep{cuddeford1993potentials,schulz2009potential}, which is feasible to the full physical space, in contrast to the disk potential based on the ring model used in most previous studies \citep{hahn2003secular,zanazzi2018planet,spalding2020stellar}, which is feasible only when planets being quite close to the disk plane. As a result, the secular full-space disk potential, $\Phi_{\mathrm{dk},i}$, is given by \citep{fu2024sculpting}
\begin{equation}
\begin{split}
	\Phi _{\mathrm{dk},i} &\approx\phi _{\mathrm{dk},i}\left\{ \frac{\pi}{8}e_{i}^{2}+\frac{\sqrt{1-e_{i}^{2}}}{S_i}\left[ S^2_i+\frac{e_{i}^{2}}{2}\left( \hat{\boldsymbol{l}}_{\mathrm{d}}\cdot \hat{\boldsymbol{e}}_i \right) ^2 \right] \right. \\
    &\quad \left. -\frac{\pi}{32}\left[ \left( 4-3e_{i}^{2} \right) S^2_i+6e_{i}^{2}\left( \hat{\boldsymbol{l}}_{\mathrm{d}}\cdot \hat{\boldsymbol{e}}_i \right) ^2 \right] \right\} .
\end{split}
\label{eq:A7}
\end{equation}
where
\begin{equation}
	\phi _{\mathrm{dk},i}=2\pi m_iG\Sigma _0a_0\
\label{eq:A8}
\end{equation}
and
\begin{equation}
	S_i=\sqrt{1-\left( \hat{\boldsymbol{l}}_{\mathrm{d}}\cdot \hat{\boldsymbol{l}}_i \right) ^2}=\sin I_{i\mathrm{d}}
\label{eq:A9}
\end{equation}
Here $I_{i\mathrm{d}}$ is the orbital inclination relative to the disk plane.

The surface density of the Metsel disk follows
\begin{equation}
\Sigma \left( r \right) =\Sigma \left( r_0 \right) \left( \frac{r_0}{r} \right) 
\label{eq:A10}
\end{equation}
where $\Sigma(r)$ is the disk surface density at radius $r$. The mass of the disk $m_{\mathrm{d}}$ is then
\begin{equation}
m_{\mathrm{d}}=\int_{r_{\mathrm{in}}}^{r_{\mathrm{out}}}{2\pi \Sigma \left( r \right) r\mathrm{d}r\simeq 2\pi \Sigma _{\mathrm{out}}r_{\mathrm{out}}^{2}}
\label{eq:A11}
\end{equation}
where $r_{\mathrm{in}}$ and $r_{\mathrm{out}}$ denote the inner and outer truncation radii of the disk. The disk evolves according to
\begin{equation}
m_{\mathrm{d}}=\frac{m_{\mathrm{d}0}}{(1+{{t}/{\tau _{\mathrm{v}}}})^{3/2}}
\label{eq:A12}
\end{equation}
where $\tau_{\mathrm{v}}$ is the timescale of the disk's dispersal.

However, the realistic evolution of the disk may not evolve homologously due to the photoevaporation effect. The disk surface density initially evolves in a homologous way driven by viscous accretion. However, when it reaches the characteristic time $\tau_{\rm{w}}$ the viscous accretion rate and photoevaporative mass loss rate become comparable at the critical photoevaporation radius $r_{\rm{c}}$. After $\tau_{\rm{w}}$ photoevaporation starves the inner disk ($r<r_{\rm{c}}$) from resupply by the outer disk’s viscous evolution, leading to a rapid dissipation of the inner disk with the viscous time $\tau _{\mathrm{v},\mathrm{in}}\ll \tau _{\mathrm{v}}$, while the outer disk ($r>r_{\rm{c}}$) continues to evolve over the viscous time $\tau_{\rm{v,out}}=\tau_{\rm{v}}$.

The evolution of the disk surface density that captures the main effect of photoevaporation can be parameterized as (assuming both the inner and outer disks follow a Mestel disk density profile)
\begin{equation}
	\Sigma \left( r,t \right) =\left\{ \begin{array}{c}
	\Sigma _{\mathrm{c}}\left( t \right) \left( {{r_{\mathrm{c}}}/{r}} \right) , r_{\mathrm{in}}\leqslant r\leqslant r_{\mathrm{c}}\\
	\Sigma _{\mathrm{out}}\left( t \right) \left( {{r_{\mathrm{out}}}/{r}} \right) , r_{\mathrm{c}}<r\le r_{\mathrm{out}}\\
\end{array} \right. 
\label{eq:A13}
\end{equation}
where
\begin{equation}
\begin{cases}
	\Sigma _{\mathrm{c}}\left( t \right) ={{\Sigma _{\mathrm{c}}\left( 0 \right)}/{\left( 1+{{t}/{\tau _{\mathrm{v},\mathrm{in}}}} \right) ^{{{3}/{2}}}}}\\
	\Sigma _{\mathrm{out}}\left( t \right) ={{\Sigma _{\mathrm{out}}\left( 0 \right)}/{\left( 1+{{t}/{\tau _{\mathrm{v},\mathrm{out}}}} \right)}}^{{{3}/{2}}}\\
\end{cases}
\\
\label{eq:A14}
\end{equation}
Here, the time $t$ is measured with respect to $\tau_{\rm{w}}$.

As a result, in the non-homologous disk evolution manner, the gravity of the disk constitutes two parts contributed by the inner and outer disks respectively. Assuming planets lie well within the inner regions (i.e., $a_i \ll r_c$), the inner disk's potential is approximately given by Eq. (\ref{eq:A7}). The outer disk's secular potential takes the form
\begin{equation}
	\Phi _{\mathrm{dkout},i}=\frac{\phi _{\mathrm{dkout},i}}{2}\left[ 5\left( \boldsymbol{e}_i\cdot \hat{\boldsymbol{l}}_{\mathrm{d}} \right) ^2-\left( \boldsymbol{l}_i\cdot \hat{\boldsymbol{l}}_{\mathrm{d}} \right) ^2+1/3-2e_{i}^{2} \right]
\label{eq:A15}
\end{equation}
where
\begin{equation}
	\phi _{\mathrm{dkout},i}=\frac{3\pi Gm_i\Sigma _{\mathrm{out}}r_{\mathrm{out}}a_{i}^{2}}{4r_{\mathrm{c}}^{2}}
\label{eq:A16}
\end{equation}

For mutual planetary interactions, due to their compact architectures, we use the Laplace-Lagrange theory to describe their orbital processions. The secular potential assuming circular orbits is given by
\begin{equation}
	\Phi _{ij}=-\frac{\phi_{ij}}{2}\left( \hat{\boldsymbol{l}}_i\cdot \hat{\boldsymbol{l}}_j \right) ^2 \,.
\label{eq:A17}
\end{equation}
Here (assuming $a_i<a_j$)
\begin{equation}
	\phi _{ij}=\frac{Gm_im_ja_i}{4a_{j}^{2}}b_{{{3}/{2}}}^{\left( 1 \right)}\left( \frac{a_i}{a_j} \right) \,,
\label{eq:A18}
\end{equation}
where $b^(1)_{3/2}$ is the Laplace coefficient, given by
\begin{equation}
    b_{{{3}/{2}}}^{\left( 1 \right)}\left( \alpha \right) =\frac{1}{\pi}\int_0^{2\pi}{\frac{\cos \theta}{\left( 1-2\alpha \cos\theta +\alpha ^2 \right) ^{{{3}/{2}}}}\mathrm{d}\theta}.
 \label{eq:A19}
 \end{equation}

This potential only captures orbital evolution of circular orbits, and in the text, this formulation is utilized for analytical purpose. In practical scenarios, initially circular orbits could be eccentricity unstable. Concerning this, we use the Gaussian ring method to model full secular gravitational interactions between planets, which is a numerical algorithm that accurately and efficiently computes the secular evolution of mutually interacting orbits. Within the mathematical framework of this method, non-resonant planetary orbits are modeled as massive wires, where each wire's line-density is inversely proportional to the planet's orbital velocity \citep{touma2009gauss}. However, this method lacks an analytical expression for $\Phi_{ij}$. In our calculations, our full secular simulation code is based on the publicly available Rings code, which implements the methodology described in \citet{touma2009gauss}.

By using the above potentials, the secular equations of motion, in terms of the vectorial elements $\boldsymbol{l}$ and $\boldsymbol{e}$, can be derived from

\begin{equation}
\begin{split}
	\dot{\boldsymbol{l}}_i=-\frac{1}{L_i}\left( \boldsymbol{l}_i\times \nabla _{\boldsymbol{l}_i}\Phi +\boldsymbol{e}_i\times \nabla _{\boldsymbol{e}_i}\Phi \right) 
	\\
	\dot{\boldsymbol{e}}_i=-\frac{1}{L_i}\left( \boldsymbol{l}_i\times \nabla _{\boldsymbol{e}_i}\Phi +\boldsymbol{l}_i\times \nabla _{\boldsymbol{e}_i}\Phi \right) 
\end{split}
\label{eq:A20}
\end{equation}

where $i={1,2,...,n}$, and $L_i=m_i\sqrt{Gm_{\star}a_i}$ is the orbital angular momentum of planet $i$.

\section{Eccentricity excitation mechanism} \label{app_EccExc}

\begin{figure}[ht!]
 \plotone{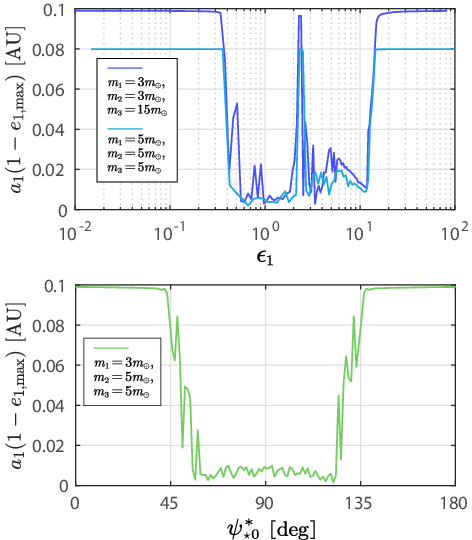}
 \caption{ Eccentricity excitation as functions of $\epsilon$ (upper panel) and $\psi_{\star0}$ (lower panel). This figure extends the analysis of Fig. \ref{fig:EccExc} of the main text of two-planet systems to the scenarios of three-planet systems. We consider a three-planet system with planetary masses $m_1=m_2=m_3=5m_{\oplus}$.
 \label{fig:EccExc2}}
 \end{figure}

 In the main text, the eccentricity excitation for two-planet systems are presented in Fig. \ref{fig:EccExc}. Through the similar method, the eccentricity excitation for three-planet systems are shown in Fig. \ref{fig:EccExc2}. 

\begin{figure}[h!]
 \plotone{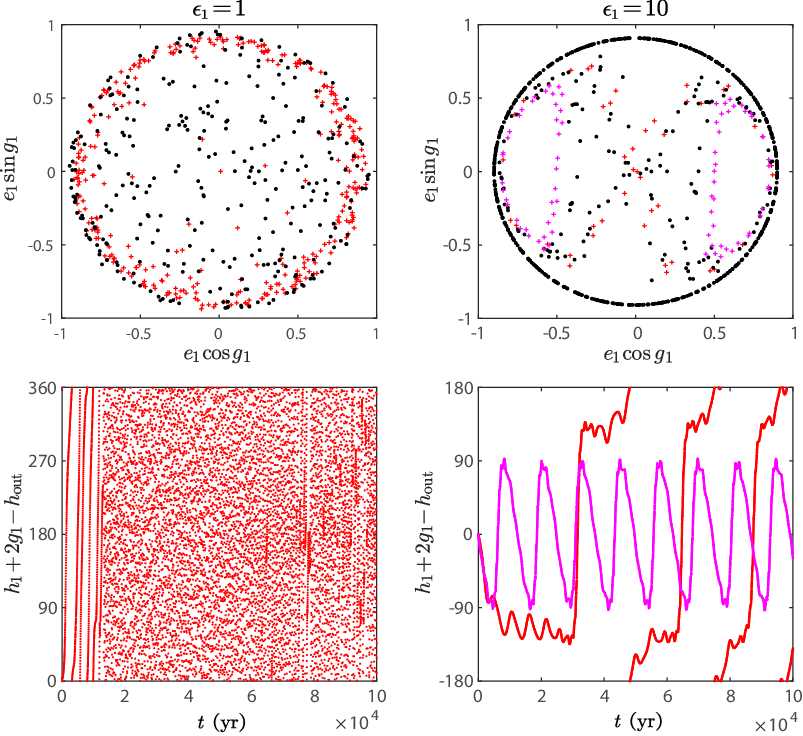}
 \caption{ Poincaré surfaces of section for chaos and a specific secular resonance identified in Fig. \ref{fig:EccExc} of the main text. In the left two panels, $\epsilon _1=1$. The section in the upper panel is defined by $h_1=180^\circ$ and $\dot{h}_1<0$, and trajectories are plotted on the $\left( e_1\cos g_1, e_1\sin g_1 \right) $ plane, where $h$ and $g$ are the longitude of ascending node and argument of perigee, respectively. The lower panel shows the the temporal evolution of the angle $h_1+2g_1-h_{\rm{out}}$ (where $h_{\rm{out}}\approx h_2$). The right two panels present results for $\epsilon _1=10$, emphasizing how eccentricity excitation is linked to a specific secular resonance involving the resonant angle $h_1+2g_1-h_{\mathrm{out}}$. The red color in these panels represents the scenario of $e_{1,0}\approx 0$ ($e_{2,0}$ and $e_{3,0}$ are set as 0). The pink color denotes the scenario where the resonant angle librates around $0^\circ$ with an amplitude of $\sim90^\circ$.
 \label{fig:PoinSec}}
 \end{figure}

 Figure \ref{fig:PoinSec} presents Poincaré surfaces of section for chaos and a specific secular resonance identified in Fig. \ref{fig:EccExc}.

\bibliography{references}{}
\bibliographystyle{aasjournal}



\end{CJK}
\end{document}